\documentclass[nofootinbib]{revtex4}
\usepackage{graphicx}
\usepackage{latexsym}
\def\be{\begin{equation}}
\def\ee{\end{equation}}
\def\bea{\begin{eqnarray}}
\def\eea{\end{eqnarray}}

\begin{document}

\title{ Scalar Gravity and Higgs Mechanism }
\author{Chi-Yi Chen${}^{a,b,c}$}
\email{chenchiyi@center.shao.ac.cn}

\author{Kang Li${}^{a}$}

\author{You-Gen Shen${}^{b,d}$}

\affiliation{${}^a$Hangzhou Teachers College, Hangzhou 310036, PR
China}
 \affiliation{${}^b$Shanghai Astronomical Observatory, Chinese
Academy of Sciences, Shanghai 200030, PR China}
\affiliation{$^{c}$Graduate School of Chinese Academy of Sciences,
Beijing 100039, China}
\affiliation{${}^d$National Astronomical
Observatories, Chinese Academy of Sciences, Beijing 100012, PR
China}

\begin{abstract}
The role that the auxiliary scalar field $\phi$ played in
Brans-Dicke cosmology is discussed. If a constant vacuum energy is
assumed to be the origin of dark energy, then the corresponding
density parameter would be a quantity varying with $\phi$; and
almost all of the fundamental components of our universe can be
unified into the dynamical equation for $\phi$. As a generalization
of Brans-Dicke theory, we propose a new gravity theory with a
complex scalar field $\varphi$ which is coupled to the cosmological
curvature scalar. Through such a coupling, the Higgs mechanism is
naturally incorporated into the evolution of the universe, and a
running density of the field vacuum energy is obtained which may
release the particle standard model from the rigorous cosmological
constant problem in some sense. Our model predicts a running mass
scale of the fundamental particles in which the gauge symmetry
breaks spontaneously. The
 running speed of the mass scale in our case could survive all
existing experiments.
\\
PACS number(s): 98.80.Cq, 12.60.Fr
\end{abstract}

\maketitle
\section{INTRODUCTION}

Recent observational data, in particular the Hubble diagram of type
I supernova \cite{perlmutter}(1998) and the fit of cosmological
parameters to the Wilkinson Microwave Anisotropy Probe (WMAP)
data\cite{WMAP}, have given support to a novel scenario for our
universe. The observable universe, which may contain three density
components, could in fact have serious departures from the
previously assumed standard cosmological model. In this novel
scenario, a dark energy dominates the universe today and drives its
acceleration. This energy must be distributed smoothly on large
scales, and be of negligible effect during early epochs. However,
the amount of dark energy may in fact be of the same order of
magnitude as the matter during a long period of cosmological
history. In theory, this problem can be resolved by making
modifications to the right-hand side of Einstein's field equation,
but may require a fine tuning of the different density components of
the universe. For example, an additional scalar field of matter may
demand a tracking behavior in playing the role of dark energy or
dark matter. Therefore, in this paper we will attempt to make
modifications to the left-hand side of Einstein's equations by
adding geometry terms.

A number of models for the dark energy have been suggested. The
model based on general relativity with a constant vacuum energy is
by far the simplest one. This is effectively the same as the
cosmological constant in standard general relativistic cosmology.
However, the presently predicted values from theory are much
greater than those inferred from observations, it is so big that
we may have to appeal to the Anthropic
Principle\cite{weiberg,garriga}. The quintessence model, which
invokes a very light and slowly evolving scalar field, requires
its potential to be so flat that it is difficult to explain how
the tiny mass of this field can stay safe from quantum
corrections. Other models include a network of topological
defects, and calls for extra dimensions. But all these have
conceptual problems which need to be further
clarified\cite{peebles,kachru,kallosh,gutperle}.

On the other hand, all searches for the signs of new physics beyond
the particle standard model have only confirmed the remarkable
success of the standard model. These confirmations have been
attributed to the success of the Higgs mechanism, but the
corresponding Higgs particle has not been found. May be we should
change the concept. As we know, the Higgs mechanism requires that
the Higgs scalar is coupled to all fundamental particles and
provides their mass. Therefore£¬it should be a universal coupling,
and possibly only gravitational interactions could do this. The
coupling between Higgs' complex scalar and the electromagnetic gauge
field should be of particular note, and for this coupling is also
required in the standard process of Higgs mechanism. As far as it is
known, there are only two kinds of interactions act on the photon
field: the electromagnetic interaction and the gravitational
interaction. Therefore if it is assumed that Higgs' complex scalar
is without charge of electricity, then the Higgs scalar can only be
interpreted as gravitational. The transfer of the gravitational
interaction may be realized through just such a coupling.
Furthermore, the particle standard model is still not able to give a
full interpretation of the origin of the hierarchy between the weak
scale and the unification scale. It may also imply that this problem
is connected with the running of the energy scale of universe.

In this letter, we would like to discuss the peculiar property of
scalar gravity. First we give a review of the cosmological property
of the Brans-Dicke's gravitational scalar filed theory in section 2,
 then we give out our scalar gravity model and discuss its
Higgs mechanism in section 3, and some conclusion remarks are given
in the last section.

\section{COSMOLOGICAL PROPERTY OF GRAVITATIONAL SCALAR FIELD}

 Brans-Dicke theory is an
alternative relativistic theory of gravity \cite{bd,b}. Compared
with general relativity, as well as the metric tensor of
space-time which describes the geometry there is an auxiliary
scalar field $\phi$ which also describes the gravity. The testing
of Brans-Dicke theory using stellar distances, the CMB temperature
and polarization anisotropy have been discussed in
\cite{gaztanaga,xueleichen}. In this section, we want to
demonstrate the dynamical property of the gravitational scalar in
the case of Brans-Dicke cosmology.

 Before applying the Brans-Dicke theory
to cosmology, we start by writing the Robertson-Walker line
element as
\begin {eqnarray}
ds^2=-dt^2+a^2(t)\tilde{g}_{ij}dx^idx^j.
\end {eqnarray}
Where i, j run from 1 to 3, $a(t)$ is the scale of the non-compact
thress-dimensional space with constant curvature $K$. The action of
Brans-Dicke theory with non-vanishing vacuum energy reads
\begin {eqnarray}
\nonumber I=&&\frac{1}{16\pi}\int d^4x\sqrt{-g}(\phi\cdot
R+\omega g^{\mu\nu}\frac{\nabla_{\mu}\phi\nabla_{\nu}\phi}{\phi})\\
&&+\int d^4x\sqrt{-g^4}\cdot({\cal{L}}_{matter}-\Lambda),
\end {eqnarray}
where $R$ is the space-time curvature scalar, $\phi$ is an
auxiliary gravitational scalar field, $\omega$ is a parameter of
Brans-Dicke theory, and the vacuum energy density from spontaneous
symmetry breaking in quantum field theory is denoted as $\Lambda$.
Here, $\Lambda>0$. Then the corresponding field equations are
\begin {eqnarray}
 \nonumber
 R_{\mu\nu}-\frac{1}{2}g_{\mu\nu}R=&&-\frac{8\pi}{\phi}T_{\mu\nu}-
\frac{1}{\phi}(g_{\mu\nu}{\phi^{;\alpha}}_{;\alpha}-\phi_{;\mu;\nu})\\
\nonumber &&-\frac{\omega}{\phi^2}\phi_{;\mu}\phi_{;\nu}
+\frac{1}{2}\frac{\omega}{\phi^2}g_{\mu\nu}\nabla_{\sigma}\phi\nabla^{\sigma}\phi\\
&&-\frac{8\pi}{\phi}\cdot\Lambda g_{\mu\nu},
\end {eqnarray}
and the field equation for $\phi$ reads
\begin {eqnarray}
\Box^2\phi={\phi^{;\mu}}_{;\mu}=\frac{8\pi}{-3+2\omega}(T_{matter}+4\Lambda).
\end {eqnarray}
 The matter
stress-energy -momentum tensor may be written as
$T_{\mu\nu}=(\rho_m+p_m)u_{\mu}u_{\nu}+p_mg_{\mu\nu}$, and then
the classically conserved perfect fluid energy momentum tensor is
\begin {eqnarray}
\frac{\partial \rho_m}{\partial
t}=-3\frac{\dot{a}}{a}(\rho_m+p_m).
\end {eqnarray}

After a straightforward calculation using Equation (1), we also
obtain the non-zero components of the Ricci tensor
\begin {eqnarray}
^{3+1}R_{00}=&&-3\frac{\ddot{a}}{a};\\
^{3+1}R_{ij}=&&(\frac{\ddot{a}}{a}+2\frac{\dot{a}^2}{a^2}+2\frac{K}{a^2})g_{ij};
\end {eqnarray}
Therefore, the fundamental equations of Brans-Dicke cosmology are
\begin {eqnarray}
1+\frac{K}{\dot{a}^2}=\frac{8\pi}{3H^2}[\frac{\rho}{\phi}-\frac{\omega}{16\pi}\frac{\dot{\phi}^2}{\phi^2}
-(\frac{3H}{8\pi}\frac{\dot{\phi}}{\phi})
+\frac{\Lambda}{\phi}];&&\\
 \nonumber \frac{\ddot{a}}{a}=-\frac{4\pi}{3}[\frac{(-6+2\omega)\rho+6\omega
 p}{\phi(-3+2\omega)}-
 4(\frac{\omega}{16\pi}\frac{\dot{\phi}^2}{\phi^2})&&\\
 -2(\frac{3H}{8\pi}\frac{\dot{\phi}}{\phi})+\frac{-6-4\omega}{-3+2\omega}(\frac{\Lambda}{\phi})].
\end {eqnarray}

As was discussed earlier, if we here assume that the vacuum energy
plays the role of dark energy in the present universe, therefore
\begin {eqnarray}
\Omega_{vacuum}=\frac{8\pi}{3H^2}(\frac{\Lambda}{\phi}),
\end {eqnarray}
 which is no longer a constant
for a fixed Hubble parameter $H$. There would be an obvious
depression in $\Omega_{vacuum}$ if the scalar field $\phi$ of
Brans-Dicke theory rolls to a large number. In addition, the
dynamical equations of $\phi$ can be derived from Equation (4), and
are
\begin {eqnarray}
\nonumber
\frac{\ddot{\phi}}{\phi}=&&-\frac{8\pi(\rho-3p)}{\phi(-3+2\omega)}+
8\pi\cdot(-\frac{3H}{8\pi}\frac{\dot{\phi}}{\phi})-\\
&&\frac{32\pi}{-3+2\omega}\cdot(\frac{\Lambda}{\phi}).
\end {eqnarray}
As a remarkable character, it is urgently necessary to note that the
Equation (11) contains almost all of cosmological density components
demonstrated by Friedman's Equations (8-9).

\section{ SCALAR GRAVITY AND HIGGS MECHANISM}
As a generalization of Brans-Dicke theory, we propose a new theory
of gravity with a complex scalar field $\varphi$ which is coupled
to the curvature scalar. It is natural to construct the action as
\begin {eqnarray}
\nonumber I=&&\frac{1}{16\pi}\int
d^4x\sqrt{-g}[\kappa\varphi\varphi^*\cdot R(t,x)+\omega
g^{\mu\nu}D_{\mu}\varphi(D_{\nu}\varphi)^*-\lambda(\varphi\varphi^*)^2]\\
&&+\int d^4x\sqrt{-g}\cdot{{\cal{L}}}_{matter}.
\end {eqnarray}
Here the coupling constants $\kappa$, $\omega$ and $\lambda$ are
all dimensionless, so it is consistent with the requirement of
renormalizability. In addition, we also restrict them to be
positive in our model.

In principle, the curvature scalar can always be decomposed into
\begin {eqnarray}
R(t,x)=\bar{R}(t)+\tilde{R}(t,x),
\end {eqnarray}
here $\bar{R}(t)$ can be regarded as the average value of the
scalar curvature:$\bar{R}(t)=<R(t,x)>$ and $\tilde{R}(t,x)$ is the
local perturbation
 induced by the actual matter distributing. In addition, it is natural to
 be assumed that in cosmology the homogeneous
isotropic component $\bar{R}(t)$ would dominate in the dynamical
equation of the universe. As far as the $RW$ metric is concerned,
this component can be written as
\begin {eqnarray}
\bar{R}(t)=6\frac{\ddot{a}}{a}+6\frac{\dot{a}^2}{a^2}+6\frac{K}{a^2}.
\end {eqnarray}
It is clear that the sign of the cosmological curvature scalar
$\bar{R}(t)$ could change with the evolution of our universe
theoretically. On the other hand, the potential of complex scalar
$\varphi$ in our case can be written as
\begin {eqnarray}
V(\varphi)=-\kappa
R(t,x)\varphi\varphi^*+\lambda(\varphi\varphi^*)^2.
\end {eqnarray}
The definition of physical vacuum may still require the space
homogeneous and isotropic. Therefore, the actual vacuum should be
considered according to the following
formula:$<V>(\varphi)=-\kappa
\bar{R}(t)\varphi\varphi^*+\lambda(\varphi\varphi^*)^2$. When
$\bar{R}(t)$ evolves
into a positive quantity, there exist non-trivial minimums. The
value of these minimums are distributed on the circle
\begin {eqnarray}
|\varphi|=\sqrt{\frac{\kappa
\bar{R}(t)}{2\lambda}}:=\frac{v}{\sqrt{2}}.
\end {eqnarray}
Hence the gauge symmetry breaks spontaneously. For convenience, we
only consider the $U(1)$ gauge symmetry in this Letter. The Higgs
mechanism requires the special gauge transformation
\begin {eqnarray}
\varphi(x)\longrightarrow\varphi'(x)=\eta(x)+\frac{v}{\sqrt{2}}; \\
A_{\mu}\longrightarrow B_{\mu}=A_{\mu}+\frac{1}{e}\nabla_{\mu}\xi(x);\\
D_{\mu}=\nabla_{\mu}+ieA_{\mu}\longrightarrow{D'}_{\mu}=\nabla_{\mu}+ieB_{\mu}.
\end {eqnarray}
After this transformation the equation (12) becomes,
\begin {eqnarray}
\nonumber I=&&\frac{1}{16\pi}\int
d^4x\sqrt{-g}[\kappa(\eta+\sqrt{\frac{\kappa \bar{R}(t)}{2\lambda}})^2\cdot \tilde{R}(t,x)+\omega\nabla_{\mu}(\eta+\sqrt{\frac{\kappa \bar{R}(t)}{2\lambda}})\nabla^{\mu}(\eta+\sqrt{\frac{\kappa \bar{R}(t)}{2\lambda}})+\eta^2\kappa \bar{R}(t)-3\eta^2\kappa \bar{R}(t)\\
\nonumber&&-2\sqrt{2\kappa\bar{
R}(t)\lambda}\eta^3-\lambda\eta^4+\omega
e^2B_{\mu}B^{\mu}(\eta+\sqrt{\frac{2\kappa
\bar{R}(t)}{\lambda}})\eta+\omega\frac{\kappa
\bar{R}(t)}{2\lambda}
e^2B_{\mu}B^{\mu}+\frac{\kappa^2 \bar{R}^2(t)}{4\lambda}]\\
 &&+\int d^4x\sqrt{-g}\cdot[{\bar{\cal{L}}}_{matter}(t)+{\tilde{\cal{L}}}_{matter}(t,x)].
\end {eqnarray}

In our framework, the running of the vacuum energy density with
the evolution of the universe is realized. We have
\begin {eqnarray}
\Lambda=\frac{1}{16\pi}\cdot\frac{\kappa^2
\bar{R}^2(t)}{4\lambda}.
\end {eqnarray}
 Here the Higgs field is a real gravitational scalar
$\eta$, which is generated by a spontaneous symmetry breaking of
the complex field $\varphi$. Therefore, the dynamical mass of the
Higgs field is $m_\eta=\sqrt{\frac{2\kappa \bar{R}(t)}{\omega}}$
and the dynamical equation for the Higgs field is
\begin {eqnarray}
\nonumber \Box^2(\eta+\sqrt{\frac{\kappa \bar{R}(t)}{2\lambda}})=&&\frac{1}{2\omega}[-4\kappa \bar{R}(t)\eta-6\sqrt{2\kappa \bar{R}(t)\lambda}\eta^2-4\lambda\eta^3+\\
&&2\omega e^2B_{\mu}B^{\mu}\eta+\omega
e^2B_{\mu}B^{\mu}\sqrt{\frac{2\kappa
\bar{R}(t)}{\lambda}}]+\frac{\kappa}{\omega}(\eta+\sqrt{\frac{\kappa
\bar{R}(t)}{2\lambda}})\cdot \tilde{R}(t,x).
\end {eqnarray}
We can imitate the discussion in Brans-Dicke theory and make a
rough estimate of the average of $\eta+\sqrt{\frac{\kappa
\bar{R}(t)}{2\lambda}}$ by computing the central potential of a
gas sphere with the cosmic mass density
$\bar{\rho}\sim10^{-29}g\cdot cm^{-3}$ and radius equal to the
apparent radius of the universe $r\sim10^{28}cm$. this gives an
average value
\begin {eqnarray}
\nonumber<\eta+\sqrt{\frac{\kappa
\bar{R}(t)}{2\lambda}}>&\sim&\frac{-4\kappa}{2\omega}(\eta+\sqrt{\frac{\kappa
\bar{R}(t)}{2\lambda}})\bar{R}(t)r^2\\
\nonumber&=&\frac{-2}{\omega(\eta+\sqrt{\frac{\kappa\bar{R}(t)}{2\lambda}})}\cdot\kappa(\eta+\sqrt{\frac{\kappa
\bar{R}(t)}{2\lambda}})^2\bar{R}(t)r^2\\
\nonumber&\sim&\frac{-2}{\omega(\eta+\sqrt{\frac{\kappa
\bar{R}(t)}{2\lambda}})}\cdot16\pi \bar{\rho}r^2\\
&\simeq&\frac{-2}{\omega<\eta+\sqrt{\frac{\kappa
\bar{R}(t)}{2\lambda}}>}\cdot16\pi\times10^{27}g\cdot cm^{-1}.
\end {eqnarray}
Note that the constant $1/G=1.35\times10^{28}g\cdot cm^{-1}$; hence,
we normalize $\eta+\sqrt{\frac{\kappa \bar{R}(t)}{2\lambda}}$ so
that
\begin {eqnarray}
\omega\cdot<\eta+\sqrt{\frac{\kappa
\bar{R}(t)}{2\lambda}}>^2\simeq \frac{1}{G}.
\end {eqnarray}
It is clear that the constant $\kappa$ is still possible to be
maintained in the magnitude order of ${\cal{O}}(1)$ in its post
Newtonian formalism.

In fact, It is not only the gauge boson obtains mass (see equation
20 for this boson mass). If it is assumed that the coupling
between fermions and this gravitational scalar $\varphi$ exists,
the fermions can also obtain mass in this
 picture. Here we consider the simple Higgs-lepton coupling
 \begin {eqnarray}
 G_e\cdot [\overline{e_R}\varphi^+(\matrix{{\nu_e}\cr{e}\cr})_L+(\matrix{{\overline{\nu_e}}\quad{\overline{e}}})_L\varphi e_R].
 \end {eqnarray}
 After the gauge symmetry breaks, the electron obtains mass
 \begin {eqnarray}
 m_e=G_e\sqrt{\frac{\kappa \bar{R}(t)}{2\lambda}}.
 \end {eqnarray}
 Therefore, the mass of fundamental particles in this scenario
 depend on the cosmological curvature scalar at spontaneous
 symmetry breaking, and are not uniquely fixed quantities any longer.
 However, as we discussed earlier, whether the gauge symmetry breaks
 or not is determined by the sign of the curvature scalar, and is
 also determined by the evolving energy scale of the universe.
 Hence, fundamental particles could not be distributed homogeneously
 on all physical energy scales in the present time. In addition,
 experiments have also shown that the mass of fundamental
 particles are stable at the present time. We think mass stability may be obtained by
 considering the average value of the cosmological scalar
 curvature $R(t,x)$, just in a similar way to which Newton's constant $G$
 can be related to the average value of the scalar field
 $\phi$ in Brans-Dicke theory.

 According to the Equation (26), it can also be investigated that the running speed of
 the electron mass in our case is given by
 \begin {eqnarray}
 \dot{m}_e=G_e\frac{1}{2}(\sqrt{\frac{\kappa
 \bar{R}(t)}{2\lambda}})^{-1}\cdot\frac{\kappa
 \dot{\bar{R}}(t)}{2\lambda}.
 \end {eqnarray}
 We recall the expression (14) of the curvature scalar in the homogeneous
isotropic universe, and further regard that $\dot{a}^2/a^2$ is still
in the same magnitude order with $\ddot{a}/a$ in the present
scenario, it is natural to be extrapolated that $\bar{R}(t)\sim
{\cal{O}}(H^2)$ and $\dot{\bar{R}}(t)\sim {\cal{O}}(H^3)$. Hence, we
can also estimate the observational effect of a variable electron
mass in the present time,
\begin {eqnarray}
\frac{\Delta{m}_e/{m}_e}{\Delta
t}\simeq\frac{\dot{m}_e}{{m}_e}=\frac{1}{2}\frac{\dot{\bar{R}}(t)}{\bar{R}(t)}\sim
{\cal{O}}(H)\sim10^{-17} sec^{-1}.
\end {eqnarray}

 On the basis of the cosmological principle, the distribution of the ordinary matter is assumed to be homogeneous
isotropic. Then the motion of the ordinary matter on the
cosmological scale is also homogeneous isotropic. Therefore, on
the cosmological scale, the gravity theory may approximately have
the formula of
\begin {eqnarray}
 I_{cos}=&&\frac{1}{16\pi}\int
d^4x\sqrt{-g}[\omega\nabla_{\mu}(\eta+\sqrt{\frac{\kappa \bar{R}(t)}{2\lambda}})\nabla^{\mu}(\eta+\sqrt{\frac{\kappa \bar{R}(t)}{2\lambda}})+\eta^2\kappa \bar{R}(t)-3\eta^2\kappa \bar{R}(t)\\
\nonumber&&-2\sqrt{2\kappa\bar{
R}(t)\lambda}\eta^3-\lambda\eta^4+\omega
e^2B_{\mu}B^{\mu}(\eta+\sqrt{\frac{2\kappa
\bar{R}(t)}{\lambda}})\eta+\omega\frac{\kappa
\bar{R}(t)}{2\lambda}
e^2B_{\mu}B^{\mu}+\frac{\kappa^2 \bar{R}^2(t)}{4\lambda}]\\
\nonumber &&+\int d^4x\sqrt{-g}\cdot{\bar{\cal{L}}}_{matter}(t).
\end {eqnarray}
The motion of cosmological scale can be ignored in present tests
of a gravity theory in the solar system. Hence in contrary to the
cosmology on the large scale, the gravity theory on the solar
scale as a local perturbation formula may be approximately taken
as
\begin {eqnarray}
I_{sol}=&&\frac{1}{16\pi}\int
d^4x\sqrt{-g}\cdot\kappa(\eta+\sqrt{\frac{\kappa
\bar{R}(t)}{2\lambda}})^2\cdot \tilde{R}(t,x)+\int
d^4x\sqrt{-g}\cdot{\tilde{\cal{L}}}_{matter}(t,x).
\end {eqnarray}

\section{CONCLUSION REMARKS}
 In this Letter, we have proposed a new form of scalar gravity theory, in which the gravitational
 complex scalar naturally
 provides a candidate Higgs field for Higgs mechanism. To demonstrate the dynamical property of
 such a gravitational scalar in cosmology, We have also
 investigated the fundamental equations of Brans-Dicke cosmology as an analog. In our scenario,
 the vacuum energy density is required to be running with
the evolution of the universe, which may release the particle
standard model from the rigorous cosmological constant problem in
some sense. Besides, a present testing-survivable running of the
mass scale of fundamental particles is also realized in our model
and may shed light on the hierarchy problem. As far as the spirit
of this letter is concerned, there are two outcomes may deserve to
be emphasized in this conclusion. One of the key ideas
 is that if a spontaneous symmetry breaking of
 the coupling between
 the curvature scalar and a gravitational scalar field occurs on the cosmological scale, some additional
  geometrical terms can be naturally introduced into the field equations. Secondly, we have argued that a gravitational
  scalar is also qualified to be a candidate for the Higgs field. Our present model is however still
  simplistic and will be clarified further in our forthcoming
  papers.

 {\bf Acknowledgments:}

 This work has been supported in part by
National Nature Science Foundation of China under grant No.
10273017, and the Foundation of Shanghai Development for Science and
Technology under grant No. 01-Jc14035. K.Li also recognize the
support from the Nature Science Foundation of Zhejiang Province
under the grant nos.M103042,M102011,and M102028.

\end{document}